\pdfoutput=1
\documentclass[aps,prab,reprint,floatfix,amsmath,amssymb,superscriptaddress,longbibliography]{revtex4-2}

\usepackage{graphicx}
\graphicspath{{figures/}}
\usepackage{xcolor}
\usepackage{amsthm}
\usepackage{upgreek}
\usepackage{float}
\definecolor{mBlue}{RGB}{51, 77, 167}
\usepackage[caption=false]{subfig}
\begin{document}

\title{Direct large-area observation of subsurface plastic activity in conditioned copper electrodes}

\author{Y.~Ashkenazy}
\email{Yinon.ash@mail.huji.ac.il}
\affiliation{Racah Institute of Physics, The Hebrew University of Jerusalem, Jerusalem 91904, Israel}
\affiliation{The Harvey M.\ Krueger Family Center for Nanoscience and Nanotechnology, The Hebrew University of Jerusalem, Jerusalem 91904, Israel}

\author{I.~Popov}
\affiliation{The Harvey M.\ Krueger Family Center for Nanoscience and Nanotechnology, The Hebrew University of Jerusalem, Jerusalem 91904, Israel}

\author{V.~M.~Bjelland}
\affiliation{CERN, European Organization for Nuclear Research, 1211 Geneva 23, Switzerland}
\affiliation{Department of Physics, NTNU, 7491 Trondheim, Norway}

\author{W.~L.~Millar}
\affiliation{CERN, European Organization for Nuclear Research, 1211 Geneva 23, Switzerland}

\author{W.~Wuensch}
\affiliation{CERN, European Organization for Nuclear Research, 1211 Geneva 23, Switzerland}

\date{\today}

\begin{abstract}
High-field conditioning is the process by which radio-frequency structures in particle accelerators and other high-gradient devices reach their operating fields, yet the underlying physical mechanism remains an open question.
Models and indirect measurements point to subsurface dislocation dynamics, but large-area structural measurements have been missing.
We present electron backscatter diffraction measurements spanning millimeter-scale regions on a copper cathode conditioned at pulsed direct-current fields up to $\sim$80~MV/m in a sloped-anode geometry, which imposes a known gradient of field exposure across a single electrode.
Across nine regions of interest spanning this exposure range, the mean intragrain misorientation of field-exposed regions exceeds that of unexposed references by $\sim$75\%; the difference is reproduced by three independent misorientation metrics and confirmed by Kolmogorov--Smirnov tests.
To our knowledge, this is the first large-area observation of structural differences between conditioned and unconditioned regions of a high-field electrode.
The misorientation separates into three tiers (high-field center and edge, low-field periphery, and unexposed reference) that match the spatial profile of the conditioning-state variable $E_S$ predicted by Monte Carlo simulations.
These observations point to the evolving subsurface dislocation population as a candidate physical basis of conditioning.
\end{abstract}

\maketitle


\section{Introduction}
\label{sec:introduction}

High-field vacuum breakdown limits the performance of particle accelerators, x-ray free-electron lasers, and other high-gradient devices~\cite{wuensch_fundamental_2026}.
\emph{Conditioning}, the progressive increase of voltage-holding capability through controlled exposure to high fields, is the primary operational means of reaching high gradients. Yet, the physical mechanism driving it remains an open question~\cite{wuensch_fundamental_2026}.
Monte Carlo simulations and experiments with spatially varying fields describe conditioning through a phenomenological state variable, $E_S$, the local field level to which each surface element has been conditioned; its microscopic origin has not been identified~\cite{bjelland_field_dependent_2026}.

Vacuum-breakdown data already point to dislocation dynamics as a limiting process~\cite{wuensch_fundamental_2026}.
Field-holding capability increases from FCC to BCC to HCP metals, the same order as the barriers to dislocation motion, linking field holding to dislocation mobility~\cite{wuensch_fundamental_2026}; metals of all three crystal structures have been high-field tested in the same pulsed direct-current system used here~\cite{serafim_effects_2025}.
A thermodynamic defect model relates the breakdown rate to the effect of the field stress on the formation enthalpy of near-surface dislocation loops~\cite{nordlund_defect_2012}.
The mobile-dislocation-density fluctuation (MDDF) model treats breakdown initiation as a critical transition in the collective motion of dislocations driven by the tensile Maxwell stress of the applied field~\cite{engelberg_stochastic_2018,engelberg_theory_2019}.
This model reproduces the observed $E^{30}$ scaling of breakdown rate with field and the temperature dependence of field-holding capability~\cite{wuensch_fundamental_2026}, and its predictions for pre-breakdown dark-current spikes are consistent with measurements~\cite{engelberg_dark_2020}.
A key prediction is that conditioning involves evolution of the near-surface dislocation structure~\cite{wuensch_fundamental_2026}.

Several experimental observations support a link between dislocation-based mechanisms and conditioning, although the evidence so far is indirect or local.
Cross-sectional transmission electron microscopy (TEM) of conditioned hard-copper cathodes has revealed dislocation-denuded zones in the top $\sim$200~nm of field-exposed regions~\cite{jacewicz_surface_2025,wuensch_fundamental_2026}, providing local evidence of field-driven near-surface plastic rearrangement.
Soft (heat-treated) copper conditions more slowly than hard (as-machined) copper~\cite{korsback_vacuum_2020}, consistent with their different initial dislocation populations, which the same cyclic load should deplete in one case and build up in the other.
At the macroscopic level, conditioning kinetics point in the same direction: the breakdown rate scales with the number of high-field pulses rather than the number of breakdowns~\cite{degiovanni_comparison_2016,wuensch_fundamental_2026}, implying that the applied field itself drives the conditioning effect rather than surface arc damage.
Despite these advances, differences in dislocation structure between conditioned and unconditioned areas have not yet been directly observed at large scales~\cite{wuensch_fundamental_2026}.

The field acts on the electrode through the Maxwell tensile stress, a surface load that the metal carries as elastic stress even though the field itself is screened at the surface.
Its magnitude, $\sigma_M = \varepsilon_0 E^2/2 \approx 0.028$~MPa at 80~MV/m, is roughly $10^3$ times smaller than the macroscopic yield stress of annealed copper ($\sim$30--70~MPa).
Nevertheless, studies of very-high-cycle fatigue have established that oscillatory stresses well below yield can produce cumulative, irreversible dislocation rearrangement when sustained over $10^{8}$--$10^{10}$ cycles~\cite{mughrabi_cyclic_2009,stanzl-tschegg_vhcf_2007}.
The cycle-count regime is therefore relevant to the $\sim$10$^{9}$ conditioning pulses applied here, while the much smaller Maxwell stress amplitude motivates the MDDF threshold-free description introduced above.

In this work, we report large-area electron backscatter diffraction (EBSD) measurements on the copper cathode from the sloped-electrode conditioning experiment of Ref.~\cite{bjelland_field_dependent_2026}, for which both the conditioning history and the spatial breakdown distribution are known.
EBSD-derived orientation gradients are established proxies for geometrically necessary dislocation (GND) density~\cite{pantleon_resolving_2008,konijnenberg_3d_ebsd_gnd_2015}, and together with local TEM cross-sections, they connect near-surface observations to millimeter-scale plastic evolution.
We compared field-exposed regions with internal reference regions on the same cathode and with a separate unexposed cathode.
Because the central comparison is made within a single specimen exposed to a continuous field gradient, it is insensitive to cathode-to-cathode differences in microstructure and composition.
The regions of interest were chosen far from individual breakdown craters, so the measured microstructure reflects the cumulative material response to field exposure rather than local arc damage.
Regions that experienced high fields show intragrain misorientation, a proxy for GND density, approximately 75\% higher than unexposed references.
To our knowledge, this is the first large-area observation of dislocation-related microstructural differences between conditioned and unconditioned regions of a high-field electrode.
The misorientation separates into three tiers of field exposure (high-field center and edge, low-field periphery, and unexposed reference) that follow the spatial profile of $E_S$ predicted by the simulations, suggesting that the evolving subsurface dislocation population is a candidate physical mechanism for conditioning.

\section{Experimental methods}
\label{sec:methods}

\subsection{Sloped-anode geometry}
\label{sec:geometry}

\begin{figure}[htbp]
    \centering
    \includegraphics[width=\linewidth]{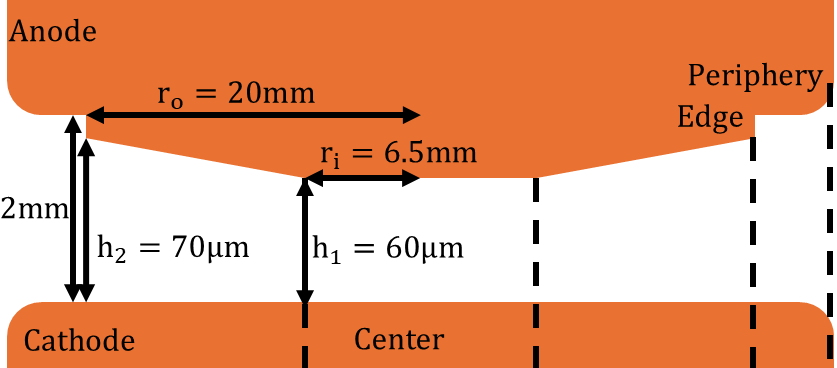}
\caption{Sloped-anode geometry.
The anode has a flat central region of radius $r_i = 6.5$~mm, separated from the cathode by a uniform gap $h_1 = 60~\upmu$m. It is surrounded by a linearly sloped annulus extending to $r_o = 20$~mm, where the gap reaches $h_2 = 70~\upmu$m.
The local surface field on the cathode, therefore, decreases monotonically with radius, defining the center, edge, and periphery regions of the field-exposed (FE) cathode used throughout the text.}
    \label{fig:electrode_geometry}
\end{figure}

The electrodes studied in this work exhibit field holding, breakdown, and conditioning similar to those of radio-frequency (RF) accelerating structures. The specific electrodes are the same sloped (frustum) electrodes whose design, conditioning, and breakdown-position analysis are reported in Ref.~\cite{bjelland_field_dependent_2026}.
Both electrodes are oxygen-free electronic (OFE) copper, vacuum heat-treated at $\sim$10$^{-6}$~mbar following a brazing-temperature cycle peaking at 835$\,^\circ$C (Sec.~\ref{sec:samples}).
The anode was machined with a frustum profile (Fig.~\ref{fig:electrode_geometry}): a flat central region of radius $r_i = 6.5$~mm separated from the cathode by a uniform gap $h_1 = 60~\upmu$m, surrounded by a linearly sloped annulus extending to the anode edge at $r_o = 20$~mm where the gap increases to $h_2 = 70~\upmu$m.
The local surface electric field is $E(r) = V_{\mathrm{applied}}/d(r)$, producing a uniform field for $r \le r_i$ and a radially decaying field for $r > r_i$.
At the maximum conditioning voltage, the peak surface field reached $\sim$80~MV/m at the center and decreased to $\sim$69~MV/m at the outer edge of the active area~\cite{bjelland_field_dependent_2026}.

The cathode was conditioned in the CERN pulsed direct-current (DC) Large Electrode System (LES) using 1~$\upmu$s pulses at a 1~kHz repetition rate.
During conditioning, breakdown positions were recorded by optical triangulation.
The resulting breakdown density varied systematically with radial position: approximately 24~breakdowns\,cm$^{-2}$ in the center (highest field), 12~breakdowns\,cm$^{-2}$ in the middle annulus, and 5~breakdowns\,cm$^{-2}$ in the outer region~\cite{bjelland_field_dependent_2026}.
This gradient, much weaker than the $E^{30}$ scaling of instantaneous breakdown rate, reflects the counteracting effect of field-driven conditioning, in which regions exposed to higher fields also condition more rapidly.

The sloped-anode geometry provides a single cathode surface with a controlled, continuously varying field-exposure history.
Measurements from the center (high field), edge (intermediate field), and periphery (negligible field) can therefore be compared on the same sample.
Results from different surface locations were also compared to a reference sample that underwent an identical preparation and heat treatment but was never exposed to high fields.
This external reference is essential because the large cathode area can exhibit location-dependent microstructure (e.g., due to machining strain and heat-flow variations).

\begin{figure}[htbp]
\centering
\subfloat[Low-magnification view of a field-exposed ROI.\label{fig:sem_lowmag}]{%
  \includegraphics[width=0.72\columnwidth]{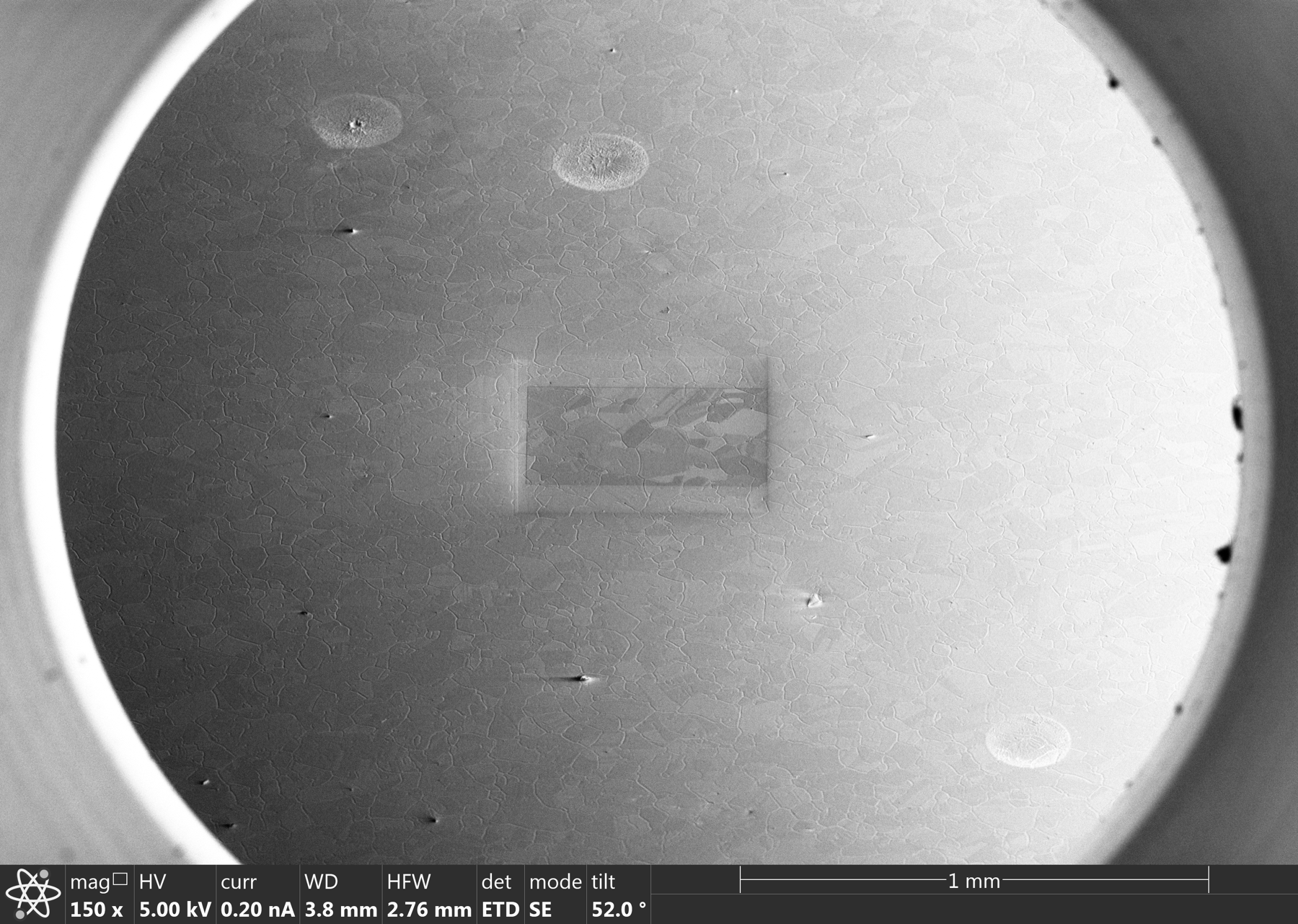}%
}\\[2pt]
\subfloat[Electron-induced.\label{fig:sem_FEroi2}]{%
  \includegraphics[width=0.49\columnwidth]{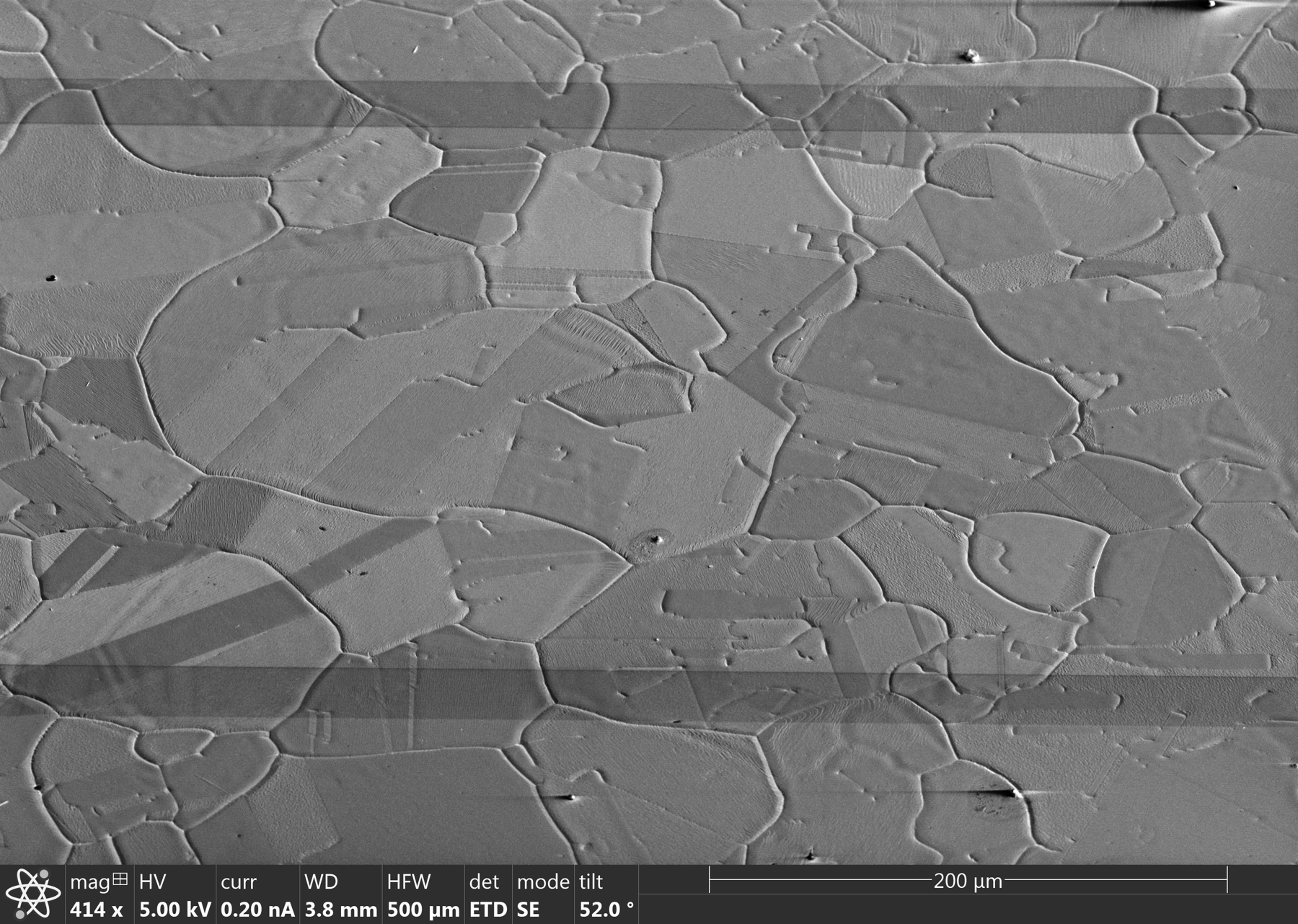}%
}\hfill
\subfloat[Ion-induced.\label{fig:sem_FEroi2_new}]{%
  \includegraphics[width=0.49\columnwidth]{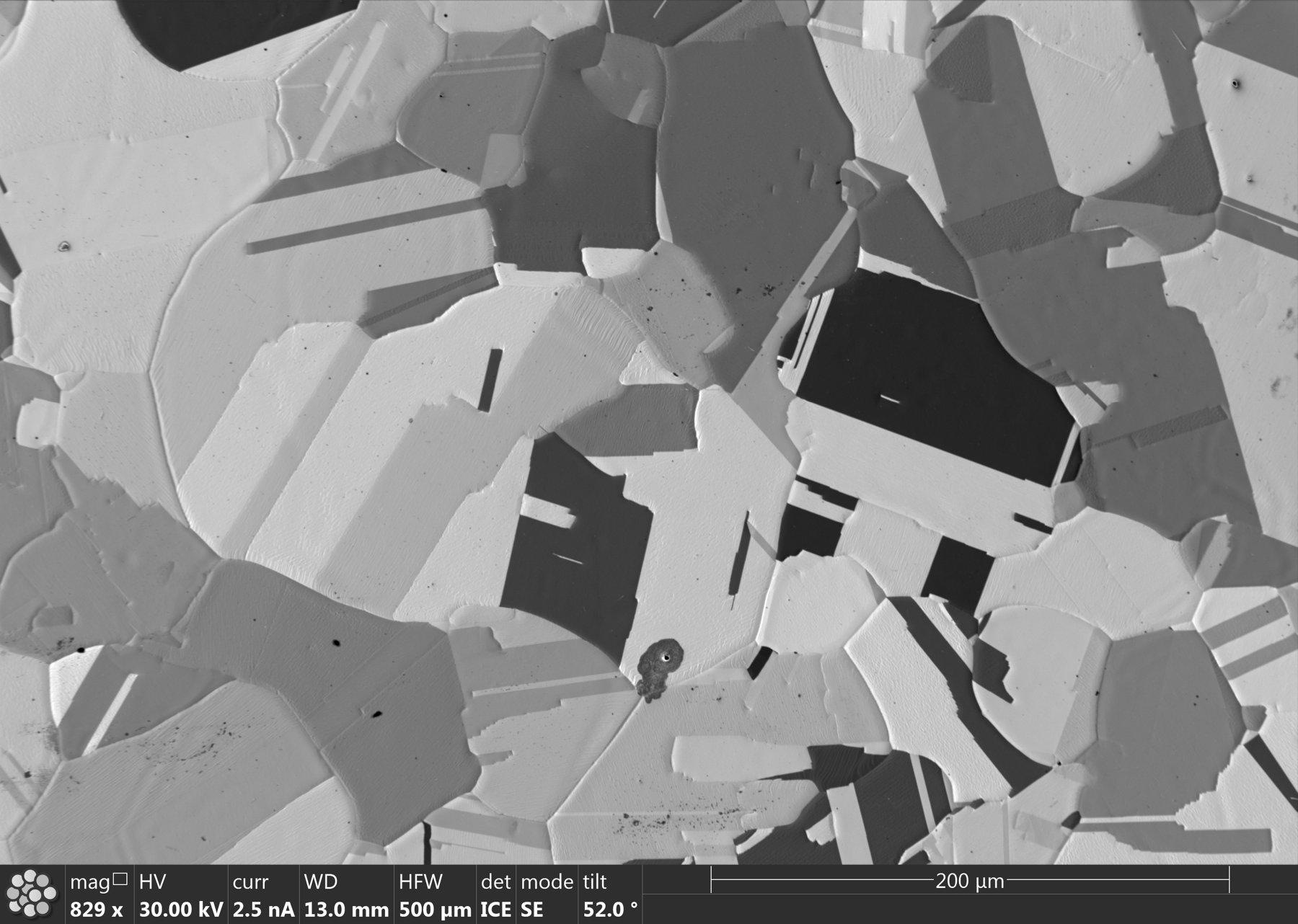}%
}
\caption{Secondary-electron images of field-exposed ROIs after FIB cleaning, before EBSD acquisition.
(a) Low-magnification view showing a cleaned ROI (central rectangle) hundreds of micrometers from the nearest breakdown craters.
(b) Electron-beam and (c) ion-beam images of a cleaned FE center ROI; the uniform grain-orientation contrast under ion imaging in (c) marks the cleaning endpoint (Sec.~\ref{sec:geometry}).}
\label{fig:sem_FE_rois}
\end{figure}

Because cross-sectional TEM provides only highly local information~\cite{jacewicz_surface_2025} that is difficult to extrapolate to the full electrode surface, we used large-area EBSD-based orientation imaging.
For each 0.5-mm-wide region of interest (ROI), we first used scanning electron microscopy (SEM) to select a region hundreds of microns away from breakdown craters (Fig.~\ref{fig:sem_FE_rois}).
This placement ensured that the measured microstructure reflected the cumulative intrinsic response to field exposure rather than local damage from individual breakdown events.
We then used a focused ion beam (FIB) to remove a thin surface oxide layer and acquired EBSD maps to characterize the grain structure and local orientation within the cleaned ROIs.

The FIB cleaning was performed using a Ga-ion Helios Nanolab 460F1 Dual Beam system (Thermo Fisher Scientific).
Rectangular ROIs measuring $550 \times 300~\upmu$m were scanned with a normally incident 30~keV, 2.5~nA Ga$^+$ beam over a $3072 \times 2048$ pixel scanning matrix at a dwell time of 500~ns--3~$\upmu$s per pixel.
The cleaning endpoint was determined in real time by monitoring the ion-induced secondary-electron image (ICE detector): scanning was stopped once uniformly strong grain-orientation contrast appeared across the entire ROI (Fig.~\ref{fig:sem_FEroi2_new}).
The resulting milling depth, measured by atomic force microscopy (Dimension Icon~XR, Bruker), did not exceed 10~nm in any of the cleaned ROIs, ensuring that the subsurface dislocation structure was probed without introducing measurable FIB damage.

\subsection{Samples and regions of interest}
\label{sec:samples}

Both cathodes were diamond-turned from the center to the periphery on OFE-grade copper substrates, resulting in a surface roughness of $R_a < 20$~nm.
They were then vacuum heat-treated at $\sim$10$^{-6}$~mbar following a brazing-temperature cycle rather than a full bonding cycle~\cite{profatilova_behaviour_2019}: a 100$\,^\circ$C/h ramp to 795$\,^\circ$C with a 6~h hold, then a 100$\,^\circ$C/h ramp to 835$\,^\circ$C with a 45~min hold, followed by natural cooling under vacuum~\cite{bjelland_field_dependent_2026}.
This treatment is expected to produce a recrystallized, equiaxed grain structure~\cite{korsback_vacuum_2020}.
Raw orientation-imaging analysis gives similar grain-size distributions in the field-exposed and reference cathodes (Sec.~\ref{sec:pole_maps}), with mean grain sizes of a few tens of micrometers and the largest grains approaching ${\sim}100~\upmu$m, consistent with prior measurements on similarly prepared substrates.
This grain size is consistent with the brazing-temperature cycle, and is well below the values produced by the higher-temperature ($\sim$1040$\,^\circ$C) bonding cycle used for full RF accelerating structures, which would substantially coarsen the grains~\cite{profatilova_behaviour_2019}.

Two copper samples were studied:
\begin{itemize}
\item \textbf{REF:} A reference copper cathode that was \emph{not} exposed to an electric field but underwent the identical machining and heat treatment as the field-exposed sample.
\item \textbf{FE:} The field-exposed copper cathode from the sloped-anode conditioning experiment described in Sec.~\ref{sec:geometry} and Ref.~\cite{bjelland_field_dependent_2026}.
\end{itemize}

\begin{figure}[htbp]
\centering
\includegraphics[width=0.95\columnwidth]{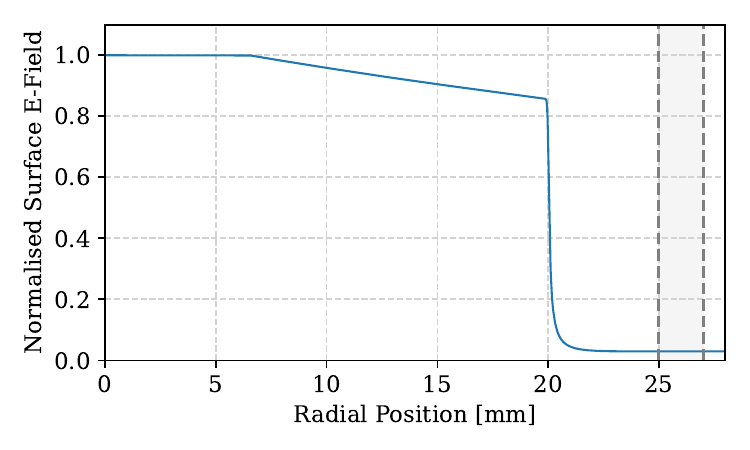}
\caption{Surface electric field on the FE cathode, normalized to the central value, as a function of radial position.
The profile was computed from a high-resolution finite-element simulation of the cathode--anode geometry.
The shaded band marks the FE Periphery ROIs ($r \approx 25$--$27$~mm), where the local field is ${\sim}2.9\%$ of the central value, corresponding to ${\lesssim}2.5$~MV/m at the maximum conditioning voltage.
The cathode edge ($r > 28$~mm), where geometric enhancement occurs, is excluded.}
\label{fig:efield_on_cathode}
\end{figure}

Multiple ROIs were examined on these samples, chosen to span the range of field exposures and breakdown densities recorded during conditioning~\cite{bjelland_field_dependent_2026}:
\begin{itemize}
\item \textbf{FE Center} ($r \le 6.5$~mm): Two ROIs (ROI1 at $r = 3.92$~mm, ROI2 at $r = 3.04$~mm) in the uniform high-field region ($\sim$80~MV/m), which experienced the highest breakdown density ($\sim$24~breakdowns\,cm$^{-2}$).
\item \textbf{FE Edge} ($r = 6.5$--$20$~mm): Two ROIs (ROI1 at $r = 9.21$~mm, ROI2 at $r = 8.65$~mm) on the sloped portion of the electrode, exposed to intermediate fields ($\sim$69--77~MV/m) and intermediate breakdown density ($\sim$12~breakdowns\,cm$^{-2}$).
\item \textbf{FE Periphery} ($r > 20$~mm): Two ROIs at the outer circumference of the FE cathode, beyond the anode radius (ROI1 at $r = 24.81$~mm, ROI2 at $r = 27.3$~mm).
A finite-element simulation of the cathode--anode geometry (Fig.~\ref{fig:efield_on_cathode}) gives a local surface field of ${\sim}2.9\%$ of the central value at this radius, corresponding to ${\lesssim}2.5$~MV/m at the maximum conditioning voltage.
These ROIs serve as an internal reference that underwent the same machining, heat treatment, and vacuum exposure as the high-field regions.
\item \textbf{REF Center:} ROI on the reference sample at $r = 6$~mm, the position that would sit below the flat anode center (zero field exposure).
\item \textbf{REF Edge:} Two ROIs on the reference sample at positions that would sit below the sloped anode edge (ROI1 at $r = 14.7$~mm, ROI2 at $r = 15.4$~mm), with zero field exposure, providing a position-matched comparison for the FE Edge ROIs.
\end{itemize}

\subsection{EBSD measurements}
\label{sec:ebsd_params}

EBSD data were acquired with an EDAX-AMETEK system (1.4-megapixel DigiView camera operated under the TEAM software) attached to an Apreo~2S SEM (Thermo Fisher Scientific).
Electron-backscatter patterns were collected at an accelerating voltage of 15~kV and a beam current of 3.2~nA, yielding an acquisition rate of approximately 40~patterns per second with a 97\% indexing success rate.
Orientation-imaging microscopy (OIM) datasets were acquired from 500~$\upmu$m-wide ROIs on a hexagonal grid with a 3~$\upmu$m step size.
Each ROI produced approximately 23{,}000--25{,}000 indexed points.
We analyzed the datasets in the as-acquired state, without applying data-cleaning filters (no wild-spike removal or confidence-index thresholding), so that the reported misorientation statistics reflect the raw measurements.

To identify plastic effects of field exposure, we used three complementary misorientation metrics: local average misorientation (LAM), the mean misorientation between each pixel and its nearest neighbors, together with local orientation spread (LOS) and kernel average misorientation (KAM), defined and compared in Sec.~\ref{sec:los_kam}.
For all three metrics, the kernel comprised the six nearest neighbors on the hexagonal grid (first-order shell, neighbor distance 3~$\upmu$m).
Pixel pairs whose misorientation exceeded 5$^\circ$ were excluded from the kernel average to suppress grain-boundary contributions~\cite{lehockey_mapping_2000,kamaya_smoothing_2010}.
The misorientation histograms reported below are accumulated over the individual nearest-neighbor pixel pairs, ${\sim}5$--$6\times10^4$ per ROI\@.

\section{Results}
\label{sec:results}

\subsection{Grain orientation maps}
\label{sec:pole_maps}

\begin{figure}[htbp]
\centering
\subfloat[Reference, center.\label{fig:pole_ref}]{%
  \includegraphics[width=0.49\columnwidth]{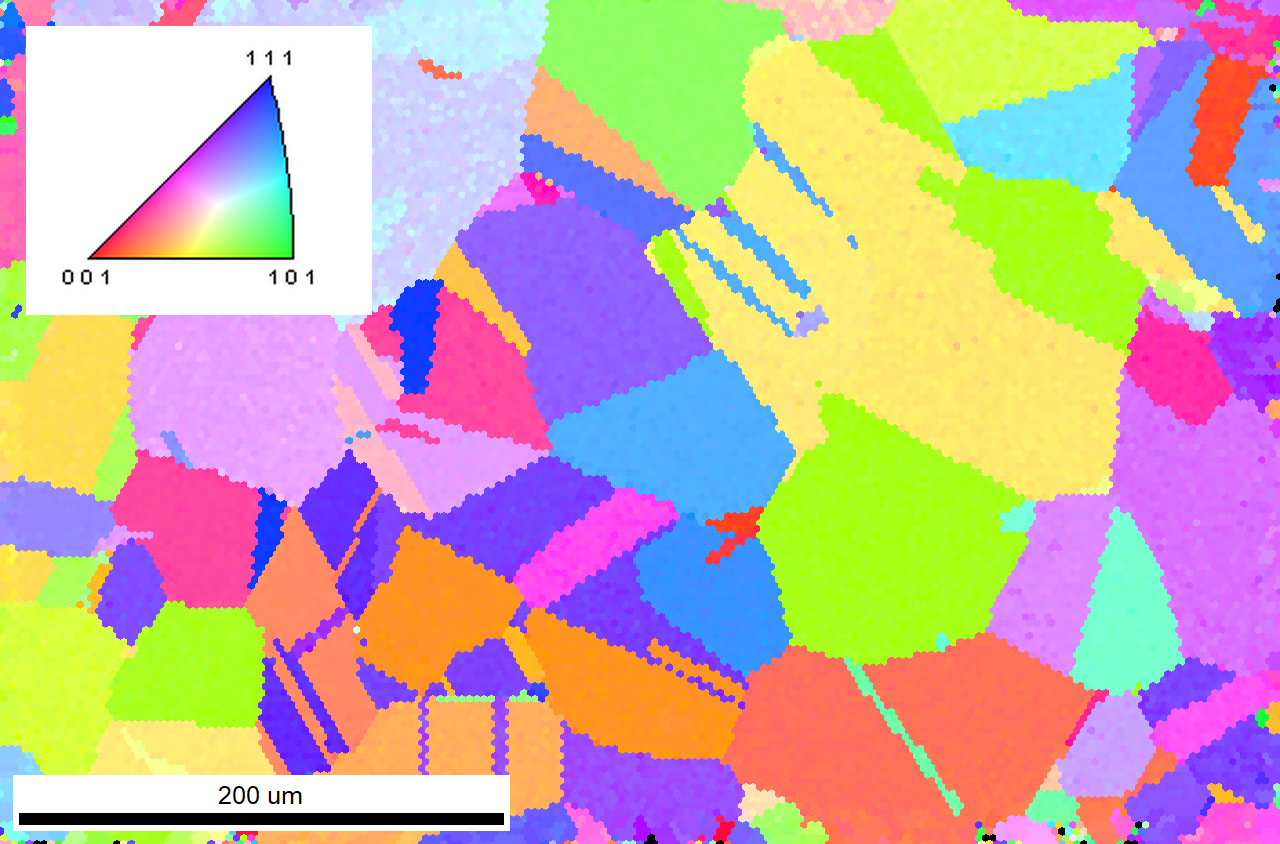}%
}\hfill
\subfloat[Field-exposed, center ROI2.\label{fig:pole_FE}]{%
  \includegraphics[width=0.49\columnwidth]{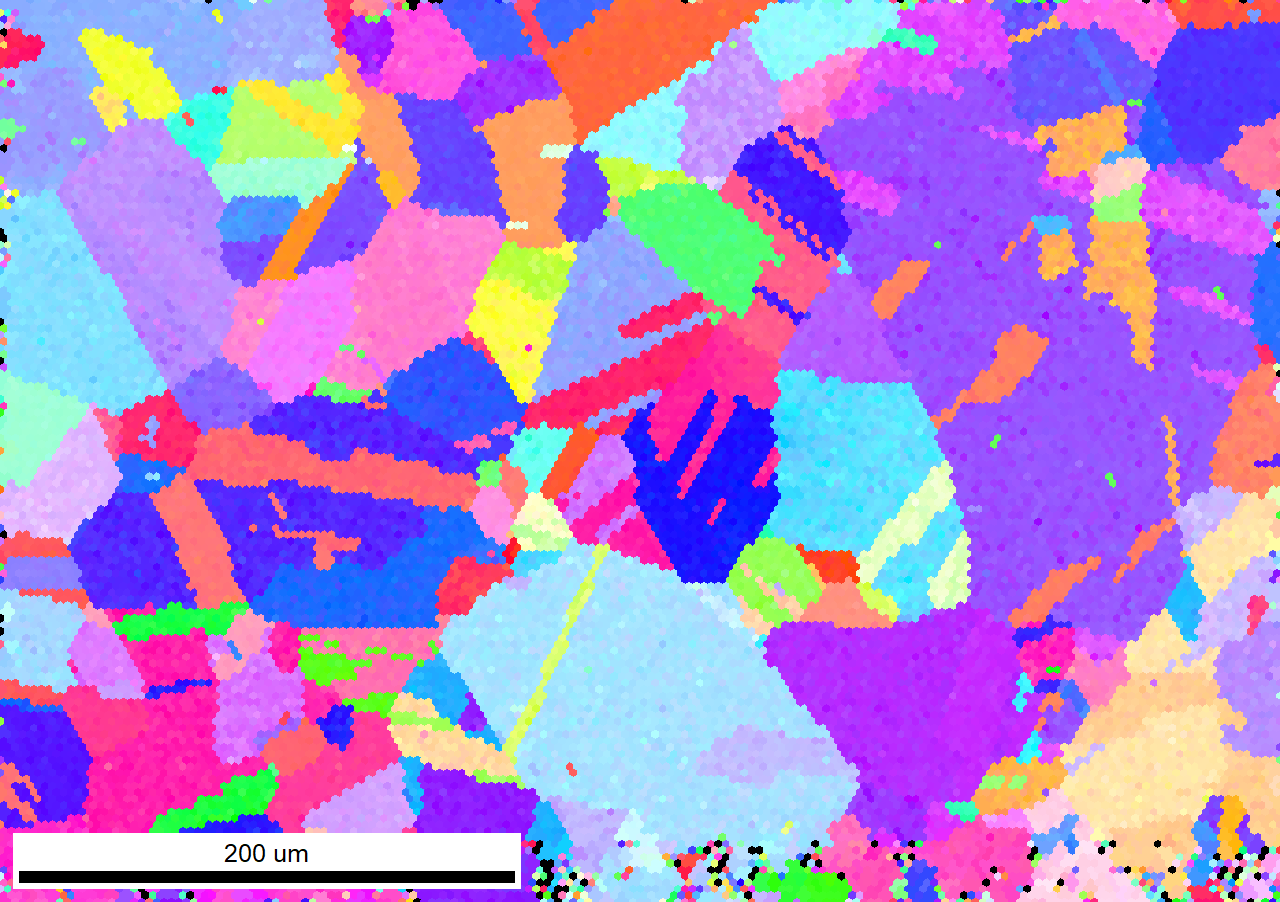}%
}
\caption{EBSD inverse pole figure maps for (a) the reference and (b) the field-exposed cathode (center ROIs); the inset triangle in (a) is the IPF color key (crystal direction along the surface normal).
Both maps show the equiaxed, randomly oriented grain structure expected after the brazing-temperature heat treatment.}
\label{fig:ebsd_pole_maps}
\end{figure}

Representative inverse pole figure (IPF) maps are shown for ROIs at the centers ($r \le 6.5$~mm) of the reference and field-exposed cathodes in Fig.~\ref{fig:ebsd_pole_maps}.

The IPF maps show a broad distribution of grain orientations in both the FE and REF samples, consistent with the recrystallized microstructure expected after the brazing-temperature heat treatment~\cite{bjelland_field_dependent_2026,korsback_vacuum_2020}.
Both cathodes were produced from the same stock and underwent identical machining and heat treatment.
Grain reconstruction from the orientation data (grain tolerance $5^\circ$, two-point minimum, with dilation to suppress spurious grains at the ROI margins; annealing twins counted as separate grains) gives mean grain diameters of $26$--$30~\upmu$m across all nine ROIs.
The reference grains are modestly larger than the field-exposed ones (${\approx}30$ versus ${\approx}26~\upmu$m on average); we ascribe this to pre-existing cathode-to-cathode microstructural variation rather than to field exposure, and it is exactly the kind of difference that the within-cathode field-gradient comparison is designed to bypass.
The misorientation hierarchy does not follow grain size: within the field-exposed cathode the mean grain size is essentially constant from center through edge to periphery, while the mean misorientation drops sharply from the high-field center and edge to the low-field periphery (Fig.~\ref{fig:mean_vs_radius}).
The misorientation differences reported below therefore arise from intragrain dislocation content rather than from differences in grain structure.

\subsection{Local average misorientation}
\label{sec:lam}

While grain sizes at the micron scale and above are not expected to change under the applied field, intragrain orientation gradients are sensitive to dislocation content.
We therefore focus on the LAM, LOS, and KAM metrics defined in Sec.~\ref{sec:ebsd_params}, which serve as proxies for GND density~\cite{pantleon_resolving_2008,konijnenberg_3d_ebsd_gnd_2015}.

\begin{figure}[htbp]
\centering
\subfloat[FE Center ROI1, wide scale (0--50$^\circ$).\label{fig:LAM_FE_center1_wide}]{%
  \includegraphics[width=0.498\columnwidth]{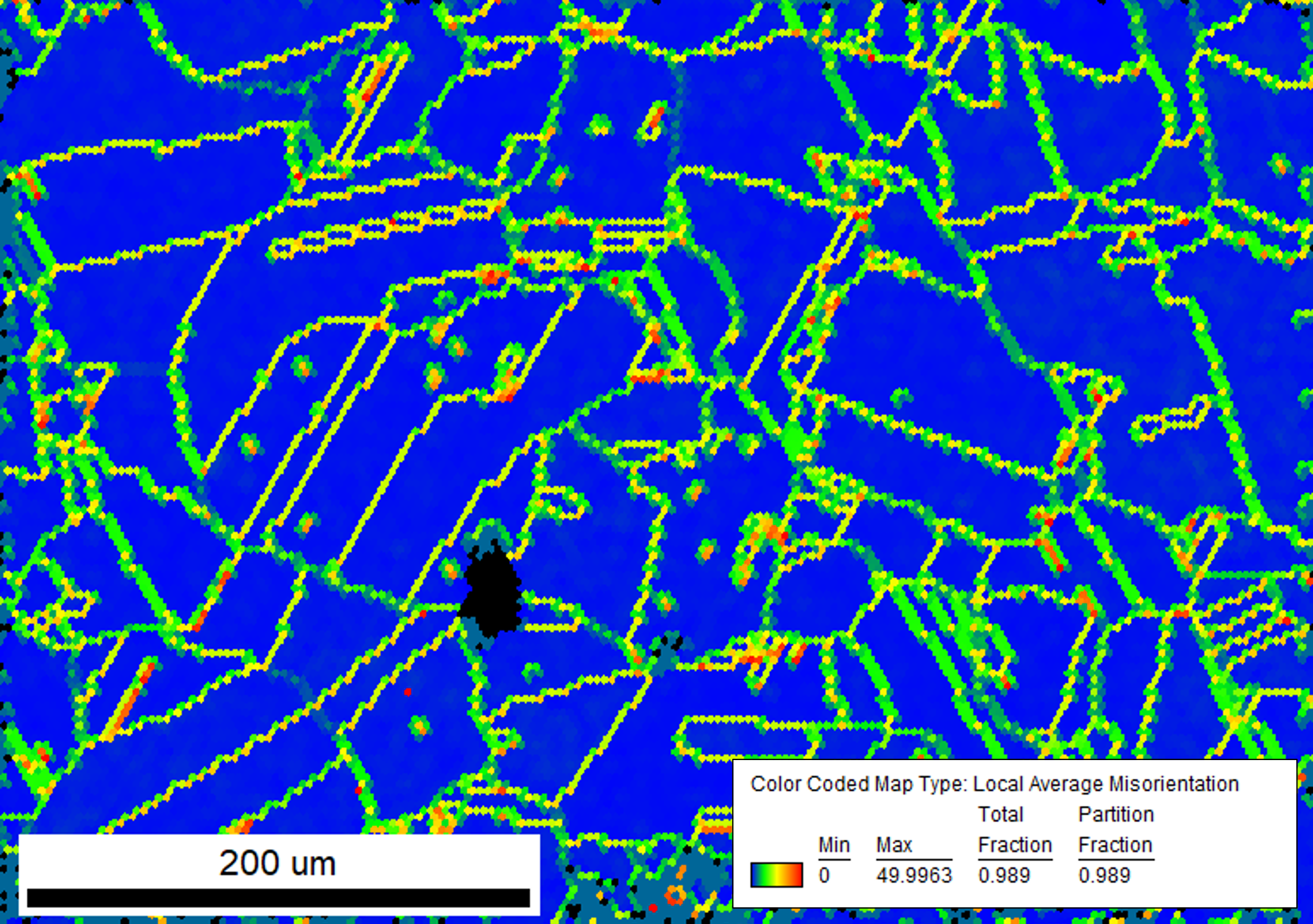}%
}\hfill
\subfloat[FE Center ROI1, low-angle scale.\label{fig:LAM_FE_center1_low}]{%
  \includegraphics[width=0.492\columnwidth]{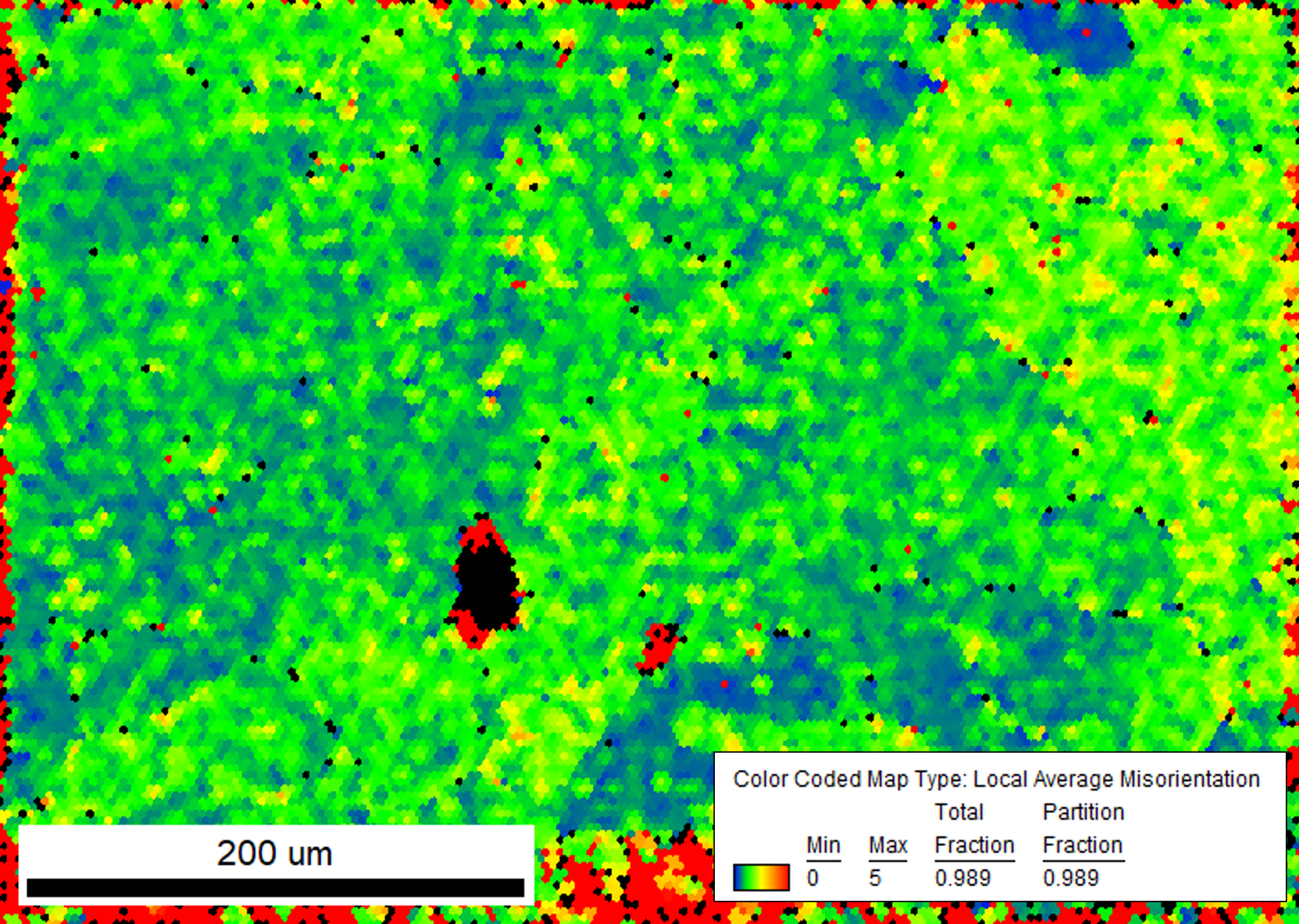}%
}
\caption{LAM with a wide (a, $0$--$50^\circ$) vs.\ low-angle (b, $0$--$5^\circ$) color scale: the wide scale is dominated by grain boundaries; the low-angle scale resolves intragrain misorientation.
All subsequent LAM, LOS, and KAM maps in this paper use the same low-angle ($0$--$5^\circ$) color scale.}
\label{fig:LAM_boundary_vs_intragranular}
\end{figure}

Figure~\ref{fig:LAM_boundary_vs_intragranular} illustrates the key distinction between grain-boundary misorientation and intragranular misorientation.
When the LAM color scale spans a wide range (0--50$^\circ$), grain boundaries dominate the contrast and intragrain variations are visually suppressed.
Restricting the scale to low angles reveals intragranular misorientation, which we use as a proxy for local dislocation density in the analysis that follows.

\begin{figure}[htbp]
\centering
\subfloat[FE Center ROI2.\label{fig:LAM_FE_center2}]{%
  \includegraphics[width=0.505\columnwidth]{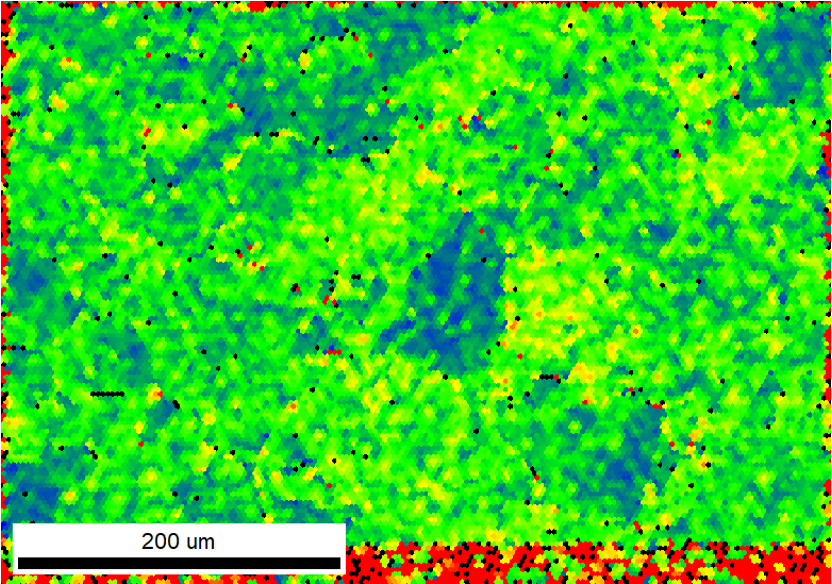}%
}\hfill
\subfloat[FE Edge ROI1.\label{fig:LAM_FE_edge1}]{%
  \includegraphics[width=0.490\columnwidth]{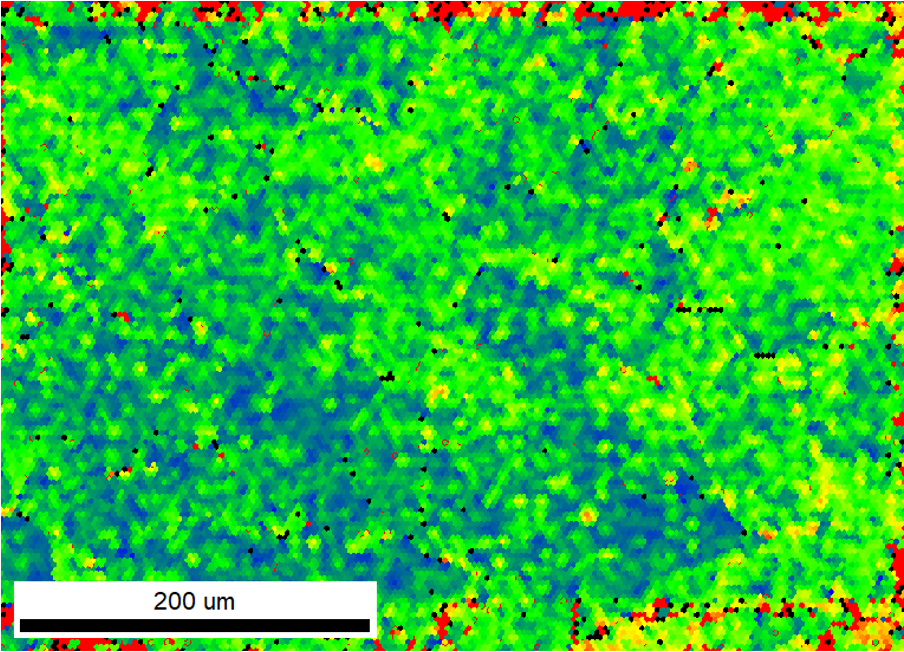}%
}
\caption{LAM maps of field-exposed ROIs at (a) the center ($r = 3.0$~mm) and (b) the edge ($r = 9.2$~mm), on the low-angle ($0$--$5^\circ$) color scale of Fig.~\ref{fig:LAM_boundary_vs_intragranular}(b).
Both regions show widespread intragrain misorientation.}
\label{fig:LAM_pair}
\end{figure}

\begin{figure}[htbp]
\centering
\subfloat[FE Edge ROI2.\label{fig:LAM_FE_Eroi2}]{%
  \includegraphics[width=0.485\columnwidth]{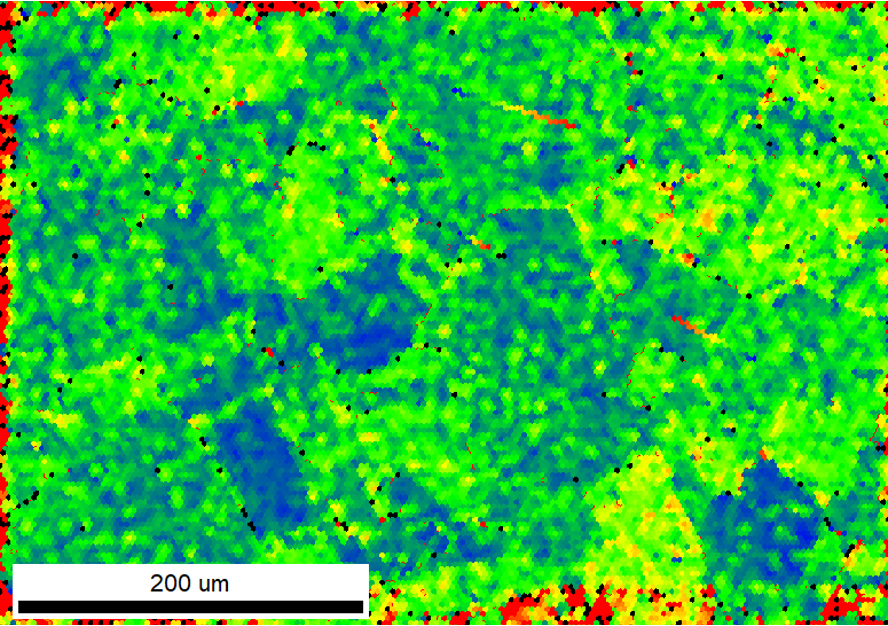}%
}\hfill
\subfloat[REF Edge ROI1.\label{fig:LAM_REF_Eroi2}]{%
  \includegraphics[width=0.510\columnwidth]{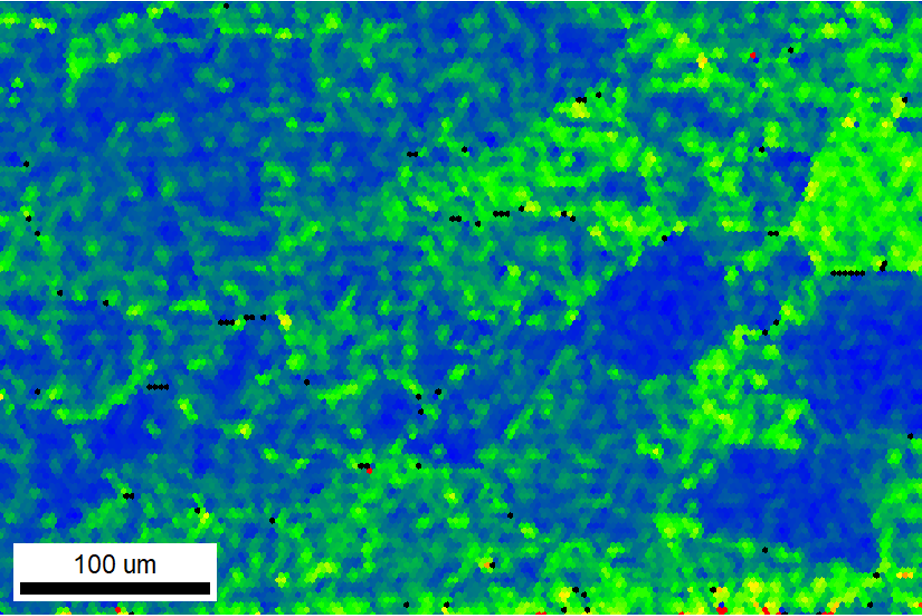}%
}
\caption{LAM maps of (a) a field-exposed edge ROI ($r = 8.7$~mm) and (b) a position-matched ROI on the unexposed reference cathode ($r = 14.7$~mm), on the same $0$--$5^\circ$ color scale.
The field-exposed region shows higher intragrain misorientation.}
\label{fig:LAM_FE_ref_pair}
\end{figure}

Using the low-angle scale, the field-exposed regions (FE center and FE edge) show elevated intragrain misorientation relative to unexposed reference regions (FE periphery and external REF sample), consistent with higher local dislocation density in the field-exposed areas (Figs.~\ref{fig:LAM_pair} and \ref{fig:LAM_FE_ref_pair}).

In contrast, unexposed reference regions exhibit substantially lower intragrain misorientation than field-exposed regions, as demonstrated in Fig.~\ref{fig:LAM_FE_ref_pair}.

\subsection{Local orientation spread and kernel average misorientation}
\label{sec:los_kam}

To confirm that the LAM trend is not metric-dependent, we computed two additional orientation-gradient measures: KAM, the mean misorientation between each pixel and its first-nearest-neighbor kernel with cross-grain pairs excluded, and LOS, the mean deviation from a kernel-averaged orientation~\cite{lehockey_mapping_2000}.
Both suppress point-to-point noise and are established proxies for local plastic strain and GND density~\cite{kamaya_smoothing_2010,zribi_ecap_dislocation_density_2019,de_vincentis_orientation_gradients_2019,field_local_orientation_gradients_2005}.
Both reproduce the LAM trend: field-exposed regions show elevated orientation gradients in all three metrics.
We therefore use LAM as the primary proxy for GND density and dislocation activity in the quantitative analysis that follows.

\subsection{Low-angle misorientation distributions}
\label{sec:histograms}

\begin{figure}[htbp]
\centering
\includegraphics[width=0.95\columnwidth]{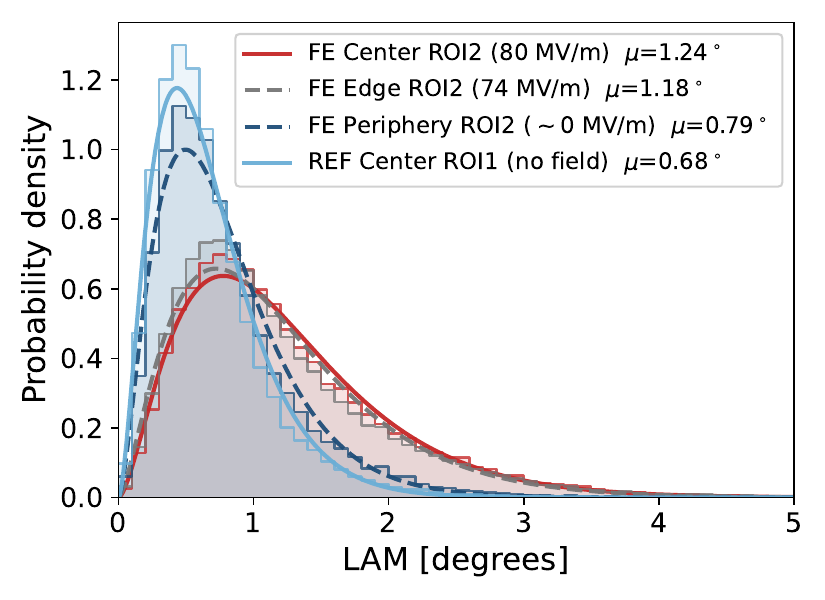}
\caption{Low-angle (0--5$^\circ$) LAM distributions, one ROI from each region (FE center, FE edge, FE periphery, and external reference).
Step histograms: 0.1$^\circ$ bins; smooth curves: interval-censored maximum-likelihood gamma fits (Table~\ref{tab:low_angle_modes}).
Field-on regions are red (FE center) and gray (FE edge); no-field regions are dark blue (FE periphery) and light blue (REF).
Center ROIs are solid, off-center ROIs dashed.
All four distributions share a common shape parameter $k \approx 2.7$; only the gamma scale $\theta$ varies with field.}
\label{fig:Lam_histograms}
\end{figure}

To quantify the variation in local misorientation, we computed low-angle (0--5$^\circ$) LAM histograms for the different ROIs (Fig.~\ref{fig:Lam_histograms}).
Regions not exposed to high fields (FE periphery and REF) consistently show lower misorientations than field-exposed regions (FE center and edge).
The histograms were fitted with gamma distributions; the extracted moments are summarized in Table~\ref{tab:low_angle_modes}.

\begin{table*}[htbp]
\caption{Low-angle (0--5$^\circ$) LAM distribution summary.
Mean and skewness are empirical moments of the 0.1$^\circ$-binned data; $k$ and $\theta$ are interval-censored maximum-likelihood gamma parameters (mean $=k\theta$); $P(\mathrm{LAM}>2^{\circ})$ is the fraction of pixels with misorientation above $2^\circ$ within the 0--5$^\circ$ window.
Uncertainties are 95\% bootstrap confidence intervals.}
\label{tab:low_angle_modes}
\begin{ruledtabular}
\begin{tabular}{l c c c c c}
\textrm{Region} & \textrm{Mean ($^\circ$)} & \textrm{$k$} &
  \textrm{$\theta$ ($^\circ$)} & \textrm{Skewness} &
  \textrm{$P(\mathrm{LAM} > 2^\circ)$} \\
\hline
FE Center ROI1 & $1.19^{+0.01}_{-0.01}$ & $2.74^{+0.03}_{-0.03}$ & $0.434^{+0.006}_{-0.006}$ & $1.36^{+0.03}_{-0.03}$ & $0.133^{+0.003}_{-0.003}$ \\
FE Center ROI2 & $1.24^{+0.01}_{-0.01}$ & $2.70^{+0.03}_{-0.03}$ & $0.458^{+0.006}_{-0.006}$ & $1.28^{+0.03}_{-0.03}$ & $0.150^{+0.003}_{-0.003}$ \\
FE Edge ROI1 & $1.21^{+0.01}_{-0.01}$ & $2.83^{+0.03}_{-0.03}$ & $0.426^{+0.006}_{-0.005}$ & $1.32^{+0.03}_{-0.03}$ & $0.139^{+0.003}_{-0.003}$ \\
FE Edge ROI2 & $1.18^{+0.01}_{-0.01}$ & $2.59^{+0.03}_{-0.03}$ & $0.457^{+0.006}_{-0.006}$ & $1.37^{+0.03}_{-0.03}$ & $0.137^{+0.003}_{-0.003}$ \\
FE Periphery ROI1 & $0.78^{+0.00}_{-0.00}$ & $2.63^{+0.03}_{-0.03}$ & $0.298^{+0.005}_{-0.004}$ & $1.84^{+0.06}_{-0.06}$ & $0.035^{+0.002}_{-0.002}$ \\
FE Periphery ROI2 & $0.79^{+0.00}_{-0.00}$ & $2.71^{+0.03}_{-0.03}$ & $0.291^{+0.004}_{-0.004}$ & $1.68^{+0.05}_{-0.05}$ & $0.033^{+0.002}_{-0.001}$ \\
REF Center ROI1 & $0.68^{+0.00}_{-0.00}$ & $2.80^{+0.04}_{-0.03}$ & $0.241^{+0.004}_{-0.004}$ & $2.05^{+0.08}_{-0.08}$ & $0.017^{+0.001}_{-0.001}$ \\
REF Edge ROI1 & $0.65^{+0.00}_{-0.00}$ & $2.92^{+0.04}_{-0.04}$ & $0.222^{+0.003}_{-0.003}$ & $2.26^{+0.12}_{-0.11}$ & $0.013^{+0.001}_{-0.001}$ \\
REF Edge ROI2 & $0.71^{+0.00}_{-0.00}$ & $2.87^{+0.04}_{-0.04}$ & $0.248^{+0.004}_{-0.004}$ & $2.10^{+0.10}_{-0.10}$ & $0.019^{+0.001}_{-0.001}$ \\
\end{tabular}
\end{ruledtabular}
\end{table*}

\begin{figure}[htbp]
\centering
\includegraphics[width=0.95\columnwidth]{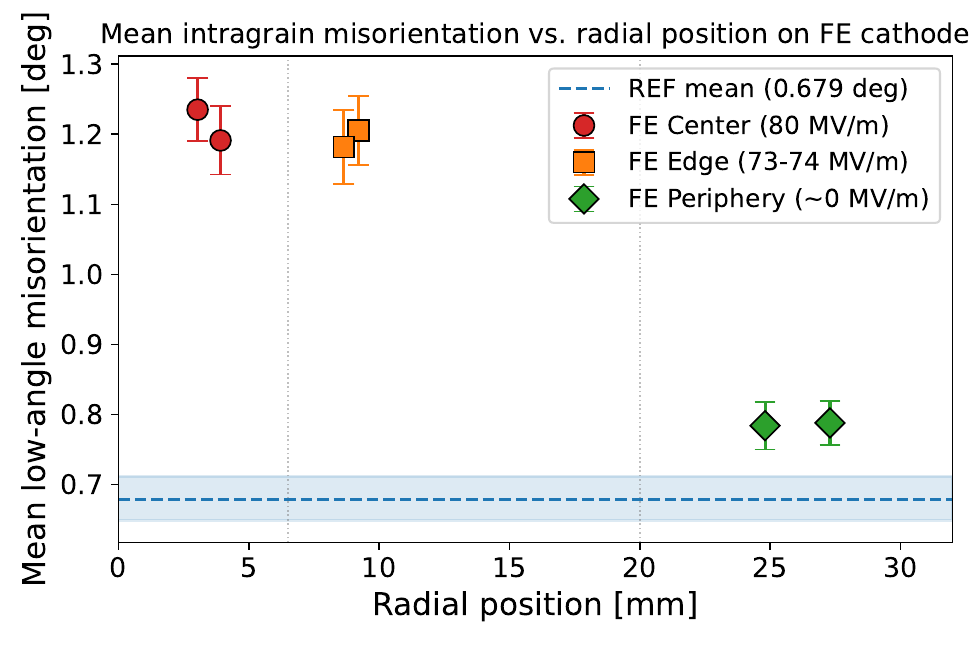}
\caption{Mean low-angle ($0$--$5^\circ$) misorientation vs.\ radial position on the field-exposed cathode.
Red circles: FE center ($\sim$80~MV/m); orange squares: FE edge ($\sim$69--77~MV/m); green diamonds: FE periphery ($\sim$0~MV/m).
Error bars are grain-corrected standard errors of the mean, $\sigma/\!\sqrt{N_{\mathrm{grains}}}$.
Dashed line and shaded band: mean and spread of the three external reference ROIs.
Dotted vertical lines at $r = 6.5$ and $20$~mm mark the uniform-gap, sloped-edge, and periphery zones.}
\label{fig:mean_vs_radius}
\end{figure}

Across all FE ROIs, the misorientation distributions have higher means and broader tails than those of the corresponding unexposed regions, whether compared to the FE periphery or to the external reference cathode (Table~\ref{tab:low_angle_modes}).
The effect is pronounced.
Beyond the ${\sim}75\%$ increase in the mean, the fraction of pixels with LAM $>2^\circ$ rises roughly eightfold, from ${\sim}0.016$ in the external reference to ${\sim}0.14$ in the field-exposed center and edge. Hence, field exposure populates the high-misorientation tail rather than merely shifting the bulk.
Two-sample Kolmogorov--Smirnov tests on the 50-bin $0$--$5^\circ$ histograms confirm that every FE center and edge distribution is distinguishable from every unexposed one ($D_{\mathrm{KS}} > 0.25$, rising above $0.31$ against the external references); the same conclusion holds under chi-squared and Anderson--Darling $k$-sample tests.
Given the ${\sim}5\times10^4$ pixel pairs per ROI, the formal $p$-values are extreme ($p < 10^{-100}$), so we characterize the difference by effect size and by the grain-corrected significance of Fig.~\ref{fig:mean_vs_radius}.
The FE periphery distributions are much closer to the external references ($D_{\mathrm{KS}} \approx 0.06$--$0.12$), though still distinguishable.
In the maximum-likelihood gamma fits, the shape parameter changes little across the ROIs: $k \approx 2.6$--$2.9$.
The scale parameter changes nearly two-fold, from $\theta \approx 0.22^\circ$ for the external references to $\theta \approx 0.46^\circ$ for the field-exposed center and edge.
Field exposure, therefore, mainly stretches the distribution toward larger misorientations.
The empirical skewness, in contrast, is not constant.
It rises monotonically with decreasing field exposure: $\sim 1.3$ at the field-exposed center and edge, $\sim 1.7$--$1.8$ at the FE periphery, and $\sim 2.0$--$2.3$ at the external reference, exceeding the gamma prediction $2/\sqrt{k}\approx 1.2$ most strongly in the reference.
External reference distributions, therefore, retain a heavier-than-gamma upper tail, while field exposure shifts the distribution toward larger misorientations.
This shift is consistent with increased dislocation activity in regions exposed to high electric fields.

Figure~\ref{fig:mean_vs_radius} summarizes the relationship between mean low-angle misorientation and radial position on the field-exposed cathode.
A three-tier hierarchy is evident: the high-field center and edge ROIs ($\sim$1.2$^\circ$) lie well above the low-field periphery ($\sim$0.79$^\circ$), which in turn exceeds the external reference ($\sim$0.68$^\circ$, shown as the horizontal band).

\begin{figure}[htbp]
\centering
\includegraphics[width=0.95\columnwidth]{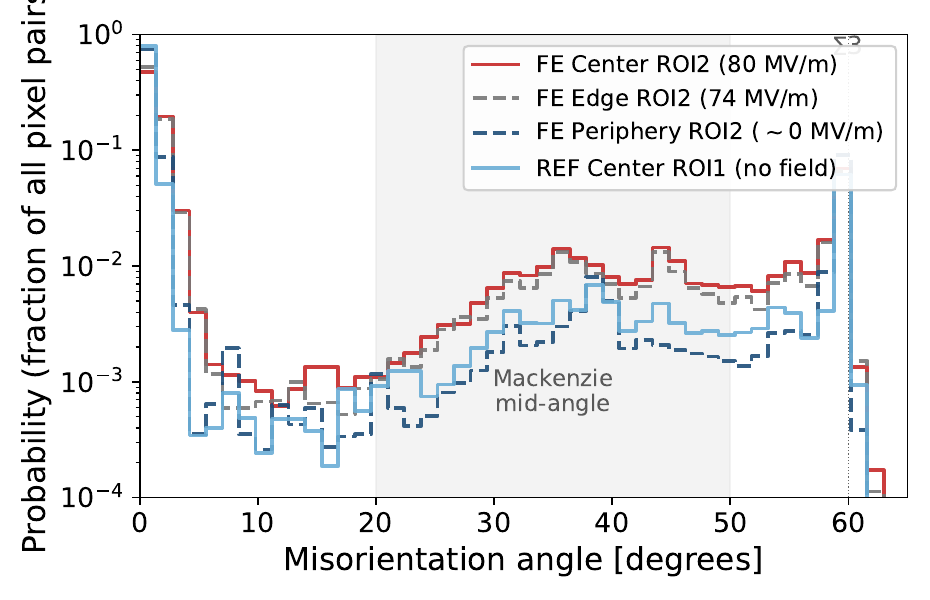}
\caption{Full-range (0--63$^\circ$) misorientation histograms for the four ROIs of Fig.~\ref{fig:Lam_histograms} (same color/linestyle), on a log probability scale (fraction of all pixel pairs).
Shaded band: Mackenzie window (20--50$^\circ$), the broad central region expected for random cubic misorientations.
Dotted line: $\Sigma 3$ twin at 60$^\circ$.
The $\Sigma 3$ peak fraction (${\sim}\,7$--$11\%$) is similar in every sample, but the pair fraction in the Mackenzie window is ${\sim}2$--$3\times$ higher in field-exposed regions, the high-angle tail of the low-angle broadening seen in Fig.~\ref{fig:Lam_histograms}.}
\label{fig:misorientation_triplet}
\end{figure}

The full-range (0--63$^\circ$) misorientation histograms (Fig.~\ref{fig:misorientation_triplet}, log probability scale) confirm that the field-induced broadening of the low-angle distribution of Fig.~\ref{fig:Lam_histograms} extends as a heavy tail into the mid-angle range 20--50$^\circ$.
This window coincides with the broad central region of the Mackenzie distribution~\cite{mackenzie_second_1958}, the misorientation-angle distribution expected between randomly oriented pairs of cubic crystals, defined on $[0^\circ, 62.8^\circ]$ with a peak near 45$^\circ$.
Physically, the window captures boundaries between subgrains that have rotated far enough to no longer count as low-angle.
Field-exposed regions populate this window at several $\times 10^{-3}$ to $10^{-2}$ per $1.4^\circ$ bin, two-to-three times higher than the unexposed regions ($\sim 2$--$5\times10^{-3}$).
By contrast, the $\Sigma 3$ coincident-site-lattice twin peak near 60$^\circ$ contains a similar pixel-pair fraction (${\sim}\,7$--$11\%$) in every sample, indicating that the annealing-twin density and the broader grain-boundary network are unaffected by field exposure.
The field-dependent broadening therefore extends from the sub-degree LAM regime into the mid-angle subgrain regime, while the high-angle twin and grain-boundary structure imposed during initial annealing is preserved.

\section{Discussion}
\label{sec:discussion}

\subsection{Misorientation and plastic deformation}
\label{sec:disc_misorientation}

The elevated intragrain misorientation in field-exposed regions, measured against both internal and external references, addresses an important open question in breakdown conditioning~\cite{wuensch_fundamental_2026}.
While the MDDF model reproduces the field- and temperature-dependent breakdown rate, and cross-sectional TEM has revealed local dislocation-denuded zones in conditioned hard copper~\cite{jacewicz_surface_2025}, differences in dislocation structure between conditioned and unconditioned areas had, until now, not been observed at large scales.
Here, mean low-angle misorientation in FE center and edge regions ($\sim$1.2$^\circ$) exceeds that in the external reference ($\sim$0.68$^\circ$) by approximately 75\%, indicating increased lattice curvature and GND density consistent with dislocation-mediated plastic processes activated by the tensile Maxwell stress~\cite{engelberg_theory_2019}.

The field itself is screened within the surface charge layer, but the load is not: the Maxwell traction acts on the surface as a mechanical boundary condition and is carried into the electrode as elastic stress, quasi-statically on the timescale of the pulse.
The electrostatic screening length therefore has no bearing on the depth of the microstructural response.

The large-scale increase in misorientation does not contradict the near-surface denuded zones of Ref.~\cite{jacewicz_surface_2025}: the two observations reflect one mechanism acting on very different initial dislocation populations.
In dense, as-machined copper the cyclic stress lets near-surface dislocations escape to the surface or annihilate in pairs, depleting the layer imaged in the TEM cross sections.
The surrounding ensemble screens the internal stress of such a layer in a Debye-like manner, over a screening length $\ell_D = 1/k_0$ with $k_0 \approx 4.2\sqrt{\rho}$ in the two-dimensional, single-slip theory of Ref.~\cite{groma_debye_2006}; for the $\rho \sim 10^{14}$~m$^{-2}$ of as-machined copper this gives $\ell_D \approx 25$~nm, so the $\sim$200~nm denuded layer is about eight screening lengths thick.
In the general geometric theory of mechanical screening this is the dipole regime, with dislocations as the screening charges~\cite{lemaitre_anomalous_2021,livne_geometric_2023}.
Neither theory, however, treats a cyclically loaded crystal with a free surface, so the thickness of the denuded layer is not predicted.
In the heat-treated copper studied here there is little to deplete; the cyclic load instead generates and rearranges dislocations.
EBSD records the resulting lattice curvature in the layer it samples, about 40~nm deep in copper at 15~kV~\cite{chen_ebsd_resolution_2011,drouin_casino_2007}; the 3-$\upmu$m step and kernel set only the lateral resolution.
This contrast is analogous to cyclic softening versus hardening under mechanical fatigue~\cite{engelberg_theory_2019,wuensch_fundamental_2026}; very-high-cycle fatigue of copper shows this directly, with annealed, low-dislocation copper cyclically hardening through dislocation accumulation while cold-drawn copper softens through dislocation rearrangement~\cite{yue_vhcf_copper_2025}.
That hard copper conditions more rapidly than soft copper~\cite{korsback_vacuum_2020} fits this picture.
Independent measurements confirm that oscillatory stress below the macroscopic yield reorganizes dislocations and leaves an EBSD-detectable signature.
Ultrasonic loading of coarse-grained nickel at a 20~MPa amplitude, with the same area imaged before and after, raised the EBSD-measured GND density by about an order of magnitude and increased the low-angle boundary fraction~\cite{zhilyaev_structure_2018} --- the same response seen here under field exposure.
The oscillatory amplitudes in such experiments (tens of MPa) far exceed the Maxwell stress; the threshold-free MDDF accumulation over ${\sim}10^9$ pulses, rather than a direct amplitude comparison, bridges this gap.
Atomistic simulations reach a similar conclusion for single-pulse loading: static fields of experimental magnitude do not by themselves trigger substantial dislocation activity on nanosecond timescales~\cite{bagchi_atomistic_2022,bagchi_precursors_2024}, so any field-driven plasticity must be cumulative.

The EBSD maps extend the earlier local TEM observations~\cite{jacewicz_surface_2025} to the full conditioned region.
Because the ROIs were placed away from crater sites, the elevated misorientation cannot be attributed to damage from individual breakdowns; it reflects the cumulative material response to high-field exposure during conditioning.
Field emission sites likewise show no spatial correlation with subsequent breakdowns~\cite{bjelland_field_dependent_2026}, consistent with conditioning being governed by material evolution below the surface rather than surface features.

The pulsed DC geometry imposes a further constraint: no RF magnetic field is present, and no pulsed surface heating occurs.
The only cyclic mechanical load is the Maxwell tensile stress, $\sigma_M = \varepsilon_0 E^2/2 \approx 0.028$~MPa at $E = 80$~MV/m, three orders of magnitude below the yield stress of the heat-treated copper used here.
In RF accelerating structures, thermal stresses from pulsed Ohmic heating can approach yield~\cite{huopana_pulsed_2006,laurent_experimental_2011} and produce grain-orientation-dependent surface roughening and slip-band formation~\cite{aicheler_evolution_2011}.
That signature is surface-localized, distinct from the subsurface misorientation reported here, and it makes the field effect difficult to isolate in RF systems.
Consistent with this distinction, the DC-conditioned surfaces studied here show no clear field-induced slip bands in scanning electron microscopy; such surface features have so far been observed only under thermal fatigue.
The DC geometry removes that thermal-loading channel, leaving Maxwell stress as the cyclic mechanical load.
This load is also distinct from classical electroplasticity, in which a high-density bulk current alters dislocation glide through the electron-wind force, or a static field acts through charge-assisted vacancy migration at high homologous temperature~\cite{conrad_electroplasticity_2000}.
Neither operates here: the cathode carries negligible bulk current, remains near ambient temperature, and screens the static field within a thin surface layer.

Beyond these loading arguments, the three-tier ordering in Fig.~\ref{fig:mean_vs_radius} (high-field center and edge above the periphery, which in turn exceeds the external reference) further argues that the misorientation increase is field-driven rather than an artifact of sample position or preparation.
The FE periphery ROIs ($r > 20$~mm) reside on the same cathode as the high-field ROIs and underwent the same machining, heat treatment, and vacuum exposure.
They therefore control for residual furnace gradients, machining strain, and vacuum history.
Yet their mean misorientation falls between the field-exposed and external reference values.
This intermediate level is consistent with the periphery having experienced a small residual fringe field during conditioning (${\lesssim}2.5$~MV/m, ${\sim}3\%$ of the central value; Fig.~\ref{fig:efield_on_cathode}), low enough to leave it closer to the pristine reference but nonzero.
The monotonic decrease of misorientation with decreasing field exposure, across both internal and external comparisons, matches the pattern expected from dose-dependent dislocation evolution rather than a spatially uniform artifact.

Several limitations bound this interpretation.
All results come from a single conditioned cathode and a single reference, so the ${\sim}75\%$ figure characterizes this material state and conditioning history rather than conditioning in general.
Field level and breakdown density vary together by construction.
Placing the ROIs far from craters removes local arc damage, but integrated exposure to distant breakdowns cannot be separated from the field exposure itself within this dataset; the observation that conditioning scales with pulse count rather than breakdown count~\cite{degiovanni_comparison_2016} favors the field as the driver.
Finally, the absolute LAM values depend on the 3-$\upmu$m step size and on the angular precision of the indexing, so the comparisons reported here are internal, between regions measured under identical acquisition and FIB-cleaning protocols.

\subsection{Connection to conditioning phenomenology}
\label{sec:disc_conditioning}

The Monte Carlo conditioning model of Ref.~\cite{bjelland_field_dependent_2026} introduces a phenomenological state variable $E_S$ that tracks the field level to which each surface element has been conditioned, and whose physical origin was left as an open question.
The spatially resolved EBSD measurements identify a possible microstructural correlate of $E_S$: intragrain misorientation, a proxy for GND density, varies with field-exposure level across the same cathode.
The same three-tier ordering in Fig.~\ref{fig:mean_vs_radius} follows the spatial profile of $E_S$ predicted by the model.
This agreement suggests that rearrangement of the subsurface dislocation population may be the physical process tracked during conditioning.
The near-constancy of the gamma shape parameter across all nine ROIs, while the scale parameter nearly doubles (Table~\ref{tab:low_angle_modes}), further supports a single underlying mechanism: field exposure changes the accumulated dislocation content, not the form of the underlying process, consistent with cumulative dose-dependent evolution.
This behavior mirrors the scaling long established for deformation-induced low-angle misorientation distributions, which collapse onto a fixed shape when scaled by their mean and depend on strain only through that mean~\cite{hughes_scaling_1998,gurao_generalized_2014}.
The low-angle misorientation produced by field exposure thus carries the same statistical signature as that produced by mechanical deformation, reinforcing its interpretation as dislocation-mediated plastic activity.

The full-range histograms (Fig.~\ref{fig:misorientation_triplet}) support the same picture.
The $\Sigma 3$ twin fraction and the high-angle boundary network are unchanged by field exposure, so the grain skeleton imposed during annealing is intact; only the intragrain and subgrain population evolves.
The elevated mid-angle population in field-exposed regions then marks subgrains that have rotated beyond the low-angle regime, the expected end state of continued GND accumulation.
The skewness trend of Table~\ref{tab:low_angle_modes}, largest in the references where the distributions retain a heavier-than-gamma tail, remains unexplained; it may reflect the as-prepared dislocation structure that field exposure progressively overwrites.

A quantitative comparison that correlates measured misorientation profiles with the radially resolved $E_S$ values from the Monte Carlo simulation will require finer-step EBSD at additional radial positions and is left for future work.

\section{Conclusions}
\label{sec:conclusions}

Field-exposed regions of a conditioned copper cathode show consistently higher subsurface intragrain misorientation than unexposed regions of the same electrode and of an identically prepared reference.
The difference is reproduced by independent misorientation metrics and confirmed by distribution-level statistical tests.
To our knowledge, this is the first large-area, spatially resolved observation of dislocation-related microstructural differences between conditioned and unconditioned regions of a high-field electrode.

The misorientation decreases stepwise with decreasing field exposure, in the same order as the spatial profile of the conditioning-state variable $E_S$ predicted by Monte Carlo simulations.
This agreement supports the picture that collective dislocation dynamics underlie conditioning and suggests that the evolving subsurface dislocation population may be the physical mechanism tracked by $E_S$.

Several directions follow.
Finer-resolution mapping would enable quantitative comparison with the predicted $E_S$ profile, and site-specific imaging could resolve the underlying dislocation structures.
The spectral content of the pulsed stress may couple preferentially to dislocation resonance frequencies, a possibility that warrants dedicated investigation.
More generally, the sub-yield response may be described as the evolution of an interacting, self-screening dislocation population, a description that would give $E_S$ a microstructural definition.
Extending the same approach to electrodes of other crystal structures, to RF-conditioned structures, and to cryogenically conditioned cathodes would test whether the subsurface response tracks dislocation mobility across materials and loading conditions.

\section*{Data Availability}
The EBSD misorientation datasets for the nine regions of interest analyzed in this work are openly available on Zenodo~\cite{ashkenazy_ebsd_data_2026}.
Other data that support the findings of this work are available from the corresponding author upon request.

\begin{acknowledgments}
This work was supported by the CERN--HUJI collaboration under Contract KE6454.
\end{acknowledgments}

\bibliography{refs}

\clearpage
\appendix
\section{Graphical summary}
\label{app:infographic}

\begin{figure*}[htbp]
  \centering
  \includegraphics[width=0.95\textwidth]{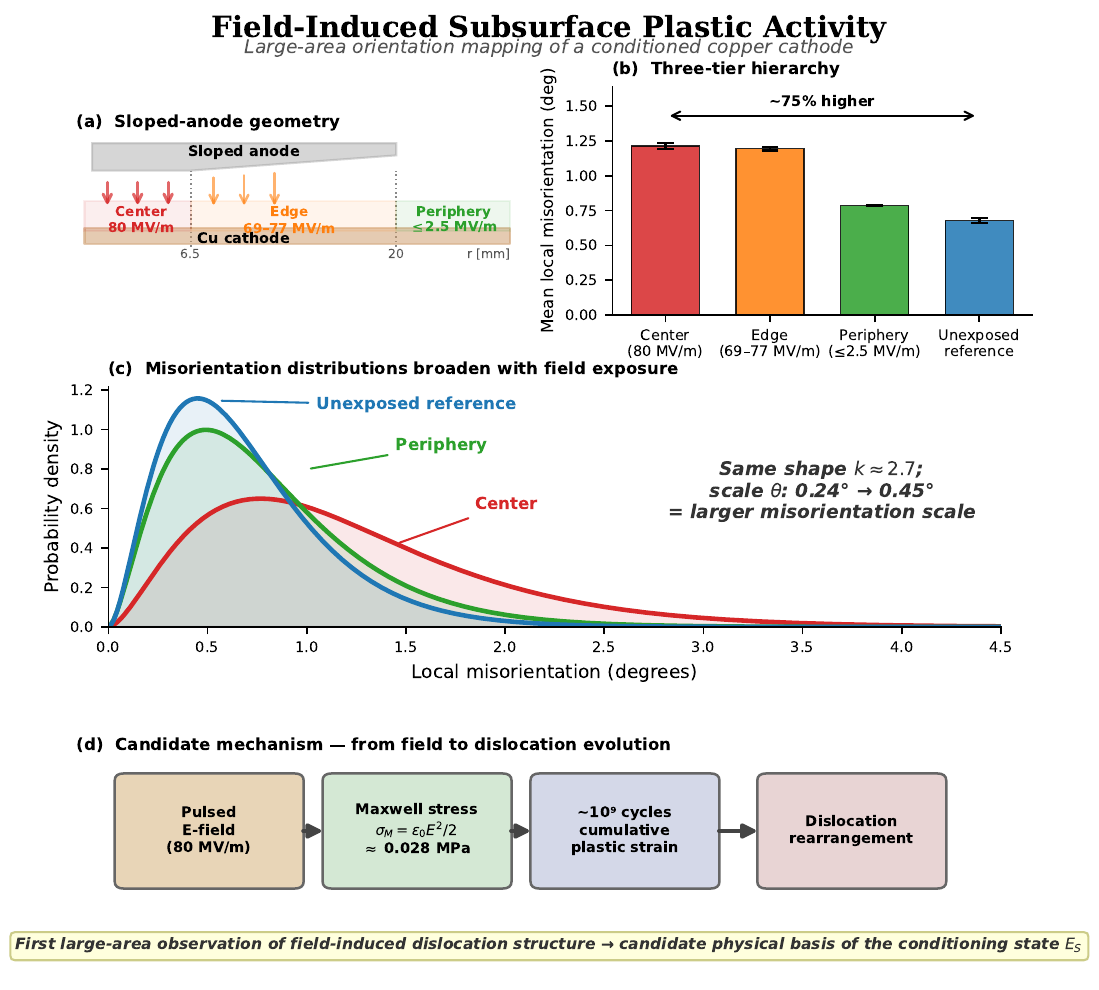}
\caption{Graphical summary.
(a)~Sloped-anode geometry: a single copper cathode spans a controlled range of surface field.
(b)~Mean local misorientation follows a three-tier hierarchy; high-field center and edge are $\sim$75\% above the unexposed reference.
(c)~Local misorientation distributions broaden with field exposure.
(d)~Candidate mechanism: cumulative Maxwell-stress-driven dislocation rearrangement, whose signature is observed here at the millimeter scale.}
  \label{fig:infographic}
\end{figure*}

\end{document}